\documentclass[12pt]{article}
\usepackage{a4wide}
\usepackage{epsfig}
\usepackage{amsmath}

\newlength{\absize}
\catcode`@=11
\def\citer{\@ifnextchar [{\@tempswatrue\@citexr}{\@tempswafalse\@citexr[]}}

\def\@citexr[#1]#2{\if@filesw\immediate
  \write\@auxout{\string\citation{#2}}\fi
  \def\@citea{}\@cite{\@for\@citeb:=#2\do
    {\@citea\def\@citea{--\penalty\@m}\@ifundefined
       {b@\@citeb}{{\bf ?}\@warning
       {Citation `\@citeb' on page \thepage \space undefined}}%
\hbox{\csname b@\@citeb\endcsname}}}{#1}}
\catcode`@=12

\begin{document}
  \thispagestyle{empty}
  \pagestyle{empty}
  \renewcommand{\thefootnote}{\fnsymbol{footnote}}
\newpage\normalsize
    \pagestyle{plain}
    \setlength{\baselineskip}{4ex}\par
    \setcounter{footnote}{0}
    \renewcommand{\thefootnote}{\arabic{footnote}}
\newcommand{\preprint}[1]{%
  \begin{flushright}
    \setlength{\baselineskip}{3ex} #1
  \end{flushright}}
\renewcommand{\title}[1]{%
  \begin{center}
    \LARGE #1
  \end{center}\par}
\renewcommand{\author}[1]{%
  \vspace{2ex}
  {\Large
   \begin{center}
     \setlength{\baselineskip}{3ex} #1 \par
   \end{center}}}
\renewcommand{\thanks}[1]{\footnote{#1}}
\begin{flushright}
%August 16, 2004
\end{flushright}
\vskip 0.5cm

\begin{center}
{\large \bf Coherent States of the Deformed Heisenberg-Weyl
Algebra in Noncommutative Space}
\end{center}
\vspace{1cm}
\begin{center}
Qi-jun Yin$^{\,1),a)}$ and  Jian-Zu Zhang$^{\,2),b),\ast)}$
\end{center}
%-----------------------------------
%   Address
%-----------------------------------
\vspace{1cm}
\begin{center}
1) Department of Physics, East China University of Science and
Technology, Shanghai 200237, P. R. China

2) Institute for Theoretical Physics, East China University of
Science and Technology, Box 316, Shanghai 200237, P. R. China
\end{center}
\vspace{1cm}
%%%%%%%%%%%%%%%%%%%%%%%%%%%%%%%%%%%%%%%%%%%%%%%%%%%%%%%%%%%%%%
\begin{abstract}
In two-dimensional space a subtle point that for the case of both
space-space and momentum-momentum noncommuting, different from the
case of only space-space noncommuting, the deformed
Heisenberg-Weyl algebra in noncommutative space is not completely
equivalent to the undeformed Heisenberg-Weyl algebra in
commutative space is clarified. It follows that there is no well
defined procedure to construct the deformed position-position
coherent state or the deformed momentum-momentum coherent state
from the undeformed position-momentum coherent state.
Identifications of the deformed position-position and deformed
momentum-momentum coherent states with the lowest energy states of
a cold Rydberg atom in special conditions and a free
particle,respectively, are demonstrated.
\end{abstract}

\clearpage

%%%%%%%%%%%%%%%%%%%%%%%%%%%%%%%%%%%%%%%%%%%%%%%%%%%%%%%%%%%%%%%%%%%%%
In recent hinting at new physics, motivated by studies of the low
energy effective theory of D-brane with a nonzero Neveu-Schwarz
$B$ field background, it shows that physics in noncommutative
space \citer{CDS,SW} is a possible candidate. Based on the
incomplete decoupling mechanism one expects that quantum mechanics
in noncommutative space (NCQM) may clarify some low energy
phenomenological consequences, and may lead to deeper
understanding of effects of spatial noncommutativity. In
literature NCQM have been studied in detail \citer{CST,JZZ04c}.
Many interesting topics of NC quantum theories have been
extensively investigated, from the Aharonov-Bohm effect to the
quantum Hall effect \citer{CDPST00,FGLMR}. Recent investigation of
the non-perturbation aspect of the deformed Heisenberg-Weyl
algebra (the NCQM algebra) \cite{JZZ04a} in noncommutative space
explored that when the state vector space of identical bosons is
constructed by generalizing one-particle quantum mechanics, in
order to maintain Bose-Einstein statistics at the non-perturbation
level described by deformed annihilation-creation operators the
consistent ansatz of commutation relations of the phase space
variables should include both space-space noncommutativity and
momentum-momentum noncommutativity. This explores some new
features of spatial noncommutativity: The spectrum of the angular
momentum of a two-dimensional system possesses fractional
eigenvalues and fractional intervals \cite{JZZ04a}; For a cold
Rydberg atom arranged in appropriate external electric and
magnetic fields, in the limits of vanishing kinetic energy and
diminishing magnetic field the unusual value of the lowest orbital
angular momentum shows a clear signal of spatial noncommutativity
\cite{JZZ04b}; Variances of a two-photon squeezed state in
different degrees of freedom correlates each other \cite{JZZ04c}.

In this paper we clarify a subtle point related to the equivalency
between the NCQM algebra in noncommutative space and the
undeformed Heisenberg-Weyl algebra in commutative space. For the
case of only space-space noncommuting, the phase space variables
of the NCQM algebra is related to the ones of the undeformed
Heisenberg-Weyl algebra by a singular-free linear transformation,
i.e. two algebras are equivalent. By a well defined procedure, the
deformed position-position coherent state in noncommutative space
can be obtained from the undeformed position-momentum coherent
state in commutative space \citer{BK}. But for the case of both
space-space and momentum-momentum noncommuting the situation is
different. The point is that there is no singular-free linear
transformation to relate phase space variables between two
algebras, i.e. two algebras are not equivalent. As is well known,
in this case three minimal uncertainties, respectively,
corresponding to the position-momentum, position-position and
momentum-momentum commutation relations are saturated by
corresponding coherent states. It only relates to the NCQM algebra
and has nothing to do with dynamics. Because of the
non-equivalency between two algebras, there is no well defined
procedure to construct the deformed position-position coherent
state or the deformed momentum-momentum coherent state from the
undeformed position-momentum coherent state. We show an example of
the deformed position-position coherent state: A cold Rydberg atom
arranged in appropriate electric and magnetic fields in the limit
of vanishing kinetic energy possesses non-trivial dynamics; Its
lowest energy state saturates the deformed position-position
uncertainty relation. An example of the deformed momentum-momentum
coherent state realized by the lowest energy state of a free
particle is briefly demonstrated.

In order to develop the NCQM formulation we need to specify the
phase space and the Hilbert space on which operators act. The
Hilbert space is consistently taken to be exactly the same as the
Hilbert space of the corresponding commutative system \citer{CST}.
As for the phase space we consider both space-space
noncommutativity (space-time noncommutativity is not considered)
and momentum-momentum noncommutativity \cite{NP,JZZ04a,DN}. In
this case the consistent NCQM algebra is as follows:
%%%%%%
\begin{equation}%1e
\label{Eq:xp} [\hat x_{i},\hat x_{j}]=i\xi^2\theta\epsilon_{ij},
\qquad [\hat x_{i},\hat p_{j}]=i\hbar\delta_{ij}, \qquad
%%%%%%
[\hat p_{i},\hat p_{j}]=i\xi^2\eta\epsilon_{ij},\;(i,j=1,2),
\end{equation}
%%%%%%
where $\theta$ and $\eta$  are the constant, frame-independent
parameters;
%%%%%%%%%%%%%%%%%%%%%%%%%%%%%%%%%%%%%%%%%%%%%%%%%%%%%%%%%%%%%%%%%%
%\footnote {\; Here we consider the intrinsic momentum-momentum
%noncommutativity. It means that the parameter $\eta$, like the
%parameter $\theta$, should be extremely small. This is guaranteed
%by the consistency condition.}
%%%%%%%%%%%%%%%%%%%%%%%%%%%%%%%%%%%%%%%%%%%%%%%%%%%%%%%%%%%%%%%%%%
$\epsilon_{ij}$ is an antisymmetric unit tensor,
$\epsilon_{12}=-\epsilon_{21}=1,$ $\epsilon_{11}=\epsilon_{22}=0$;
%%%%%%
$\xi=(1+\theta\eta/4\hbar^2)^{-1/2}$ is the scaling factor.
%%%%%%
When $\eta=0,$ we have $\xi=1$. The NCQM algebra (\ref{Eq:xp})
reduces to the one of only space-space noncommuting.

We consider a Rydberg atom with mass $\mu$ in the following
electric and magnetic fields \cite{Baxt,JZZ96,JZZ04b}: The
electric field $\vec{E}$ acts radially in the $x-y$ plane, $E_i=
-\mathcal{E}\hat x_i,$ $(i=1, 2),$ where $\mathcal{E}$ is a
constant, and the constant magnetic field $\vec{B}$ aligns the $z$
axis. The motion is constrained in the $x-y$ plane and has
rotational symmetry. The Rydberg atom is treated as a
structureless dipole moment. In reality it has the internal atomic
structure. For the following discussions effects of the internal
structure are extremely small and hence can be forgotten. The
Hamiltonian of such a Rydberg atom is (henceforth, summation
convention is used):
\begin{eqnarray}%2e
\label{Eq:Ryd-H} \hat H_{Ryd}=\frac{1}{2\mu}(\hat
p_i+\frac{1}{2}g\epsilon_{ij}\hat x_j)^2 +\frac{1}{2}\kappa\hat
x_i^2 =\frac{1}{2\mu}\hat p_i^2 +\frac{1}{2\mu}g\epsilon_{ij}\hat
p_i\hat x_j+\frac{1}{2}\mu\omega^2\hat x_i^2,
\end{eqnarray}
where the co-ordinates $\hat x_i$ refer to the laboratory frame of
the Rydberg atom. The constant parameters $g=2qB/c$ and
$\kappa=2q\mathcal{E},$ $q(>0)$ is dipole's electric charge. The
term $g\epsilon_{ij}\hat p_i\hat x_j/2\mu$ takes the Chern-Simons
interaction. The frequency $\omega=\left[g^2/4\mu^2+\kappa/\mu
\right]^{1/2},$ where the dispersive ``mass" term $g/2\mu$ comes
from the presence of the Chern-Simons term.

The NCQM algebra (\ref{Eq:xp}) changes the boson algebra of
deformed annihilation-creation operators ($\hat a_i,\hat
a_j^\dagger$) which are related to deformed phase space variables
$(\hat x_{i},\hat p_{j})$. For the Rydberg system (\ref{Eq:Ryd-H})
the deformed annihilation operator $\hat a_i$ is defined as:
\begin{equation}%3e
\label{Eq:aa+1} \hat a_i=\sqrt{\frac{\mu\omega}{2\hbar}}\left
(\hat x_i +\frac{i}{\mu\omega}\hat p_i\right).
\end{equation}
When the state vector space of identical bosons is constructed by
generalizing one-particle quantum mechanics, the maintenance of
Bose-Einstein statistics  at the deformed level of $\hat a_i$
($[\hat a_i,\hat a_j]\equiv 0$) leads to a consistency condition
\cite{JZZ04a}
\begin{equation}%4e
\label{Eq:dd} \eta=\mu^2\omega^2 \theta,
\end{equation}
and the deformed Boson algebra of $\hat a_i$ and $\hat
a_j^\dagger$ reads
\begin{equation}%5e
\label{Eq:[a,a+]1} [\hat a_1,\hat a_1^\dagger]=[\hat a_2,\hat
a_2^\dagger]=1, [\hat a_1,\hat a_2]=0;\quad [\hat a_1,\hat
a_2^\dagger] =i\xi^2\mu\omega \theta/\hbar.
\end{equation}
The first three equations in (\ref{Eq:[a,a+]1}) are the same boson
algebra as the one in commutative space. Thus Eq.~(\ref{Eq:aa+1})
is a correct definition of the deformed annihilation operator.

The last equation in (\ref{Eq:[a,a+]1}) is a new one which
correlates $\hat a_i$ and $\hat a_j^\dagger$ in deferent degrees
of freedom, codes effects of spatial noncommutativity and has some
influence on dynamics \citer{JZZ04a,JZZ04c}. It is worth noting
that it is consistent with {\it all} principles of quantum
mechanics and Bose-Einstein statistics.

If momentum-momentum is commuting, $\eta= 0$, we could not obtain
$[\hat a_i,\hat a_j]=0$. It is clear that in order to maintain
Bose-Einstein statistics for identical bosons at the deformed
level described by $\hat a_i$ and $\hat a_i^\dagger$ we should
consider both space-space and momentum-momentum
noncommutativities.

The NCQM algebra (\ref{Eq:xp}) has different possible perturbation
realizations \cite{NP}. Here we consider the following consistent
ansatz of the perturbation expansions of $\hat x_{i}$ and $\hat
p_{i}$
%%%%%%
\begin{equation}%6e
\label{Eq:hat-x-p} \hat
x_{i}=\xi(x_{i}-\frac{1}{2\hbar}\theta\epsilon_{ij}p_{j}), \quad
%%%%%%
\hat p_{i}=\xi(p_{i}+\frac{1}{2\hbar}\eta\epsilon_{ij}x_{j}).
\end{equation}
where $x_{i}$ and $p_{i}$ satisfy the undeformed Heisenberg-Weyl
algebra $ [x_{i},x_{j}]=[p_{i},p_{j}]=0,
[x_{i},p_{j}]=i\hbar\delta_{ij}.$ It is worth noting that the
determinant $\mathcal{R}_s$ of the transformation matrix $R_s$
between $(\hat x_1,\hat x_2,\hat p_1,\hat p_2)$ and
$(x_1,x_2,p_1,p_2)$ is
%%%%%%
$\mathcal{R}_{s}=\xi^4(1-\theta\eta/4\hbar^2)^2$.
%%%%%%
When $\theta\eta=4\hbar^2$, the matrix $R_s$ is singular. Thus the
NCQM algebra (\ref{Eq:xp}) and the undeformed Heisenberg-Weyl
algebra are not completely equivalent.
%%%%%%%%%%%%%%%%%%%%%%%%%%%%%%%%%%%%%%%%%%%%%%%%%%%%%%%%%%%%%%%%%%%
\footnote{\; For the case of only space-space noncommuting,
$\eta=0$, the situation is different.  The determinant
$\mathcal{R}_{uns}$ of the transformation matrix $R_{uns}$ between
$(\hat x_1,\hat x_2,\hat p_1,\hat p_2)$ and $(x_1,x_2,p_1,p_2)$ is
$\mathcal{R}_{uns}\equiv 1$ which is singular-free. Thus for the
case of only space-space noncommuting the NCQM algebra
(\ref{Eq:xp}) with $\eta=0$ and the undeformed Heisenberg-Weyl
algebra are equivalent.}
%%%%%%%%%%%%%%%%%%%%%%%%%%%%%%%%%%%%%%%%%%%%%%%%%%%%%%%%%%%%%%%%%%
Eq.~(\ref{Eq:hat-x-p}) should be correctly explained as
perturbation expansions of $\hat x_{i}$ and $\hat p_{i}$.

The perturbation expansions of $\hat a_i$ and $\hat a_{i}^\dagger$
are as follows
\begin{equation} %7e
\label{Eq:hat-a-a1} \hat
a_{i}=\xi\left(a_{i}+
\frac{i}{2\hbar}\mu\omega\theta\epsilon_{ij}a_j\right),\quad
%%%%%%%%%%%%%%%%%%%%%%%%%%%%%%%%%%%%%%%%%%%%%%%%%%%%%%%%%%%%%%%%%%
\hat a_{i}^\dagger=\xi\left(a_{i}^\dagger
-\frac{i}{2\hbar}\mu\omega\theta\epsilon_{ij}a_j\right),
\end{equation}
where $a_i$ and $a_j^\dagger$ satisfy the undeformed boson algebra
$[a_i, a_j^\dagger]=\delta_{ij}, [a_i,a_j]=0$.
Eq.~(\ref{Eq:hat-a-a1}) are consistent with the NCQM algebra
(\ref{Eq:xp}) and (\ref{Eq:hat-x-p}). The determinant
$\mathcal{R^{\; \prime}}_s$ of the transformation matrix $R^{\;
\prime}_s$ between $(\hat a_1,\hat a_2,\hat a^\dagger_1,\hat
a^\dagger_2)$ and $(a_1,a_2,a^\dagger_1,a^\dagger_2)$ is also
singular at $\theta\eta=4\hbar^2$. Eq.~(\ref{Eq:hat-a-a1}) should
be correctly explained as perturbation expansions of $\hat a_{i}$
and $\hat a^\dagger_{j}$.

In the following we study dynamics of a cold Rydberg atom
described by Eq.~(\ref{Eq:Ryd-H}). This system is exactly
solvable. But here we are interested in the limiting case of
vanishing kinetic energy. In this limit the Hamiltonian
(\ref{Eq:Ryd-H}) shows non-trivial dynamics. First we identify the
limit of vanishing kinetic energy in the Hamiltonian formulation
with the limit of the mass $\mu\to 0$  in the Lagrangian
formulation. In the limit of vanishing kinetic energy,
$\frac{1}{2\mu}\left(\hat p_i+\frac{1}{2}g\epsilon_{ij} \hat
x_j\right)^2=\frac{1}{2}\mu \dot{\hat {x_i}}\dot{\hat {x_i}}\to
0,$ the Hamiltonian (\ref{Eq:Ryd-H}) reduces to
$H_0=\frac{1}{2}\kappa \hat x_i \hat x_i.$ The Lagrangian
corresponding to the Hamiltonian (\ref{Eq:Ryd-H}) is
%%%%%%
$L_{Ryd}= \frac{1}{2}\mu \dot{\hat {x_i}}\dot{\hat
{x_i}}-\frac{1}{2}g\epsilon_{ij}\dot{\hat {x_i}}
\hat{x_j}-\frac{1}{2}\kappa\hat x_i\hat x_i$.
%%%%%%
In the limit of $\mu\to 0$ this Lagrangian %(\ref{Eq:CS-L})
reduces to $L_0=\frac{1}{2}g\epsilon_{ij}\hat x_i\dot{\hat
{x_j}}-\frac{1}{2}\kappa \hat x_i \hat x_i.$ From $L_0$ the
corresponding canonical momentum is $\hat p_{0i}=\partial
L_0/\partial \dot{\hat {x_i}}=\frac{1}{2}g\epsilon_{ji}\hat x_j,$
and the corresponding Hamiltonian is $H_0^{\prime}=p_{0i}\dot{\hat
{x_i}}-L_0=\frac{1}{2}\kappa \hat x_i \hat x_i=H_0.$ Thus we
identify the two limiting processes. It is worth noting that when
the potential is velocity dependent, the limit of vanishing
kinetic energy in the Hamiltonian does not corresponds to the
limit of vanishing velocity in the Lagrangian. If the velocity
approached zero in the Lagrangian there would be no dynamics. The
Hamiltonian (\ref{Eq:Ryd-H}) and its massless limit have been
studied by Dunne, Jackiw and Trugenberger \cite{DJT1}.

The first equation of (\ref{Eq:Ryd-H}) shows that in the limit
$E_k\to 0$ there are constraints
%%%%%%%%%%%%%%%%%%%%%%%%%%%%%%%%%%%%%%%%%%%%%%%%%%%%%%%%%%%%%%
\footnote {\;In this example the symplectic method \cite{FJ} leads
to the same results as the Dirac method for constrained
quantization, and the representation of the symplectic method is
much streamlined.}
%%%%%%%%%%%%%%%%%%%%%%%%%%%%%%%%%%%%%%%%%%%%%%%%%%%%%%%%%%%%%%
\begin{equation}%8e
\label{Eq:Ci} C_i=\hat p_i+\frac{1}{2}g\epsilon_{ij} \hat x_j=0,
\end{equation}
which should be carefully considered \cite{MZ}. Poisson brackets
of these constraints are
%%%%%%
$\{C_i,C_j\}_P=g\epsilon_{ij}\ne 0,$
%%%%%%
so that the corresponding Dirac brackets of canonical variables
$\hat x_i$ and $\hat p_j$ can be determined, $\{\hat x_1,\hat
p_1\}_D=\{\hat x_2,\hat p_2\}_D=1/2,\; \{\hat x_1,\hat
x_2\}_D=-1/g,\;\{\hat p_1,\hat p_2\}_D=-g/4$. The Dirac brackets
of $C_i$ with any variables $\hat x_i$ or $\hat p_j$ are zero, so
that the constraints (\ref{Eq:Ci}) are strong conditions and can
be used to eliminate the dependent variables. For example, if we
choose $\hat x_1$ and $\hat p_1$ as the independent variables,
from (\ref{Eq:Ci}) we obtain $\hat x_2=-2\hat p_1/g,\;\hat
p_2=g\hat x_1/2.$ But for our purpose in the following we choose
$\hat x_1$ and $\hat x_2$ as the independent variables. From the
perturbation expansion (\ref{Eq:hat-x-p}) it follows that
\begin{equation}%9e
\label{Eq:CS-H01} H_0 =\frac{1}{2}\kappa\hat x_i\hat
x_i=\frac{1}{2\mu^{\ast}}p_i
p_i+\frac{1}{2}\mu^{\ast}\omega^{\ast2} x_i x_i
+\omega^{\ast}\epsilon_{ij}p_i x_j,
\end{equation}
%%%%%%
where the effective mass $\mu^{\ast}\equiv
4\hbar^2/\xi^2\kappa\theta^2$, and the effective frequency
$\omega^{\ast}\equiv \xi^2\kappa|\theta|/2\hbar.$ The term
$\omega^{\ast}\epsilon_{ij}p_i x_j$ is the induced Chern-Simons
interaction.

In order to solve Eq.~(\ref{Eq:CS-H01}) we define the
``coordinate" and the ``momentum" $(X, P)$ and the
annihilation-creation operators $(A, A^\dagger)$ as follows
\cite{Baxt,JZZ96}
\begin{eqnarray}%10e
\label{Eq:X1}  X=\frac{1}{2}\sqrt{\mu^{\ast}}x_1
-\frac{1}{2\omega^{\ast}}\sqrt{\frac{1}{\mu^{\ast}}}p_2,\quad
P=\sqrt{\frac{1}{\mu^{\ast}}}p_1+\omega^{\ast}\sqrt{\mu^{\ast}}x_2,
\end{eqnarray}
%%%%%%%%%%%%%%%%%%%%%%%%%%%%%%%%%%%%%%%%%%%%%%%%%%%%
\begin{eqnarray}%11e
\label{Eq:AA}  A=\frac{i}{2}\sqrt{\frac{1}{\omega^{\ast}}} P
+\sqrt{\omega^{\ast}} X, \quad
A^{\dagger}=-\frac{i}{2}\sqrt{\frac{1}{\omega^{\ast}}} P
+\sqrt{\omega^{\ast}} X.
\end{eqnarray}
Where $X$ and $P$ satisfy $\left[X,P\right]=i\hbar,$ and $A$ and
$A^{\dagger}$ satisfy $\left[A,A^{\dagger}\right]=1.$ The number
operator $N=A^{\dagger}A$ has eigenvalues $n=0,1,2,\cdots.$ The
Hamiltonian (\ref{Eq:CS-H01}) is rewritten in the form of a
harmonic oscillator of the unit mass and the frequency
$2\omega^{\ast},$
%\begin{equation}
%\label{Eq:CS-H02}
%%%%%%
$H_0= 2\omega^{\ast}\hbar\left(A^\dagger A+\frac{1}{2}\right)$.
%%%%%%
%\end{equation}
The zero-point energy
\begin{equation}%12e
\label{Eq:E0}
E_0=\omega^{\ast}\hbar=\frac{1}{2}\xi^2\kappa|\theta|.
\end{equation}
This zero-point energy can be understood on the basis of the
position-position noncommutativity (\ref{Eq:xp}) and the
corresponding deformed $\hat x-\hat x$ minimum uncertainty
relation. From Eq.~(\ref{Eq:xp}) it follows that the deformed
$\hat x-\hat x$ uncertainty relation reads
%%%%%%
$\Delta {\hat x_1} \Delta {\hat x_2}\ge \frac{1}{2}\xi^2|\theta|$.
%%%%%%
Here for any normalized state $\psi,$ $\Delta {\hat F}\equiv
[(\psi,(\hat F-\bar{\hat F})^2 \psi)]^{1/2}$, $\bar{\hat F}\equiv
(\psi,\hat F\psi)$. Taking
%%%%%%
$(\Delta {\hat x_1})_{min}=\left(\Delta {\hat
x_2})_{min}=(\frac{1}{2}\xi^2|\theta|\right)^{1/2}$
%%%%%%
it follows that the minimal energy $(\Delta E)_{min}$
corresponding to $(\Delta {\hat x_i})_{min}$ is
%\begin{equation}
%\label{Eq:dE1}
$(\Delta E)_{min} = \frac{1}{2}\kappa\left[(\Delta {\hat
x_1})_{min}^2+(\Delta {\hat
x_2})_{min}^2\right]=\frac{1}{2}\xi^2\kappa|\theta|$.
%\end{equation}
This shows $(\Delta E)_{min}=E_0.$ From this result we conclude
that the deformed $\hat x-\hat x$ coherent state is realized by
the lowest energy state of the cold Rydberg atom described by
Eq.~(\ref{Eq:Ryd-H}) in the limiting case of vanishing kinetic
energy.

%%%%%%
According to Eq.~(\ref{Eq:hat-x-p}) the perturbation expansion of
the kinetic energy term $\frac{1}{2\mu}\hat p_i^2$ leads to a
perturbation induced Chern-Simons interaction, i. e. a term like
%\begin{equation*}
$\epsilon_{ij}p_i x_j$.
%\end{equation*}
The existence of this term is a general characteristics of the
NCQM algebra (\ref{Eq:xp}). This term plays essential role in
dynamics.
%%%%%%%%%%%%%%%%%%%%%%%%%%%%%%%%%%%%%%%%%%%%%%%%%%%%%%%%%%%%%%
\footnote {\; Physical systems confined to a space-time of less
than four dimensions show a variety of interesting properties.
There are well-known examples, such as the quantum Hall effect,
high $T_c$ superconductivity, cosmic string in planar gravity,
etc. In many of these cases the Chern-Simons interaction, which
exists in 2+1 dimensions and is associated with the topologically
massive gauge fields, plays a crucial role.}
%%%%%%%%%%%%%%%%%%%%%%%%%%%%%%%%%%%%%%%%%%%%%%%%%%%%%%%%%%%%%%
From this observation we show that the deformed $\hat p-\hat p$
coherent state is realized, as an example, by the lowest energy
state of a free particle. From Eq.~(\ref{Eq:hat-x-p}) it follows
that the perturbation expansion of the Hamiltonian of a free
particle $\hat H_{free}(\hat x,\hat p)=\frac{1}{2\mu}\hat p_i\hat
p_i $ reads
%\begin{equation*}
%\label{Eq:Hf}
$\hat H_{free}(\hat x,\hat p)=\frac{1}{2\tilde{\mu}}p_i
p_i+\frac{1}{2}\tilde{\mu}\tilde{\omega}^2 x_i x_i
+\tilde{\omega}\epsilon_{ij}p_i x_j$,
%\nonumber
%\end{equation*}
where the effective mass $\tilde{\mu}\equiv \xi^{-2}\mu$ and the
effective frequency $\tilde{\omega}\equiv \xi^2|\eta|/2\mu\hbar$.
In the above equation
%Eq.~(\ref{Eq:Hf})
there are an effective harmonic potential
$\frac{1}{2}\tilde{\mu}\tilde{\omega}^2 x_i x_i$ and an effective
Chern-Simons interaction $\tilde{\omega}\epsilon_{ij}p_i x_j$.
This means that a ``free" particle in noncommutative space is not
free; it moves in the above effective potentials. Based on this
result we may guess that the noncommutativity of space originates
from some {\it intrinsic} background fields. By a similar
procedure of solving Eq.~(\ref{Eq:CS-H01}) we obtain
%%%%%%
$\hat H_{free}= 2\tilde{\omega}\hbar\left(\tilde A^\dagger \tilde
A+\frac{1}{2}\right)$,
%%%%%%
where $\tilde A$ and $\tilde A^\dagger$ are defined by a similar
equation (\ref{Eq:AA}), in which ($X$, $P$) and ($\mu^{\ast}$,
$\omega^{\ast}$) are replaced, respectively, by ($\tilde X$,
$\tilde P$) and ($\tilde {\mu}$, $\tilde {\omega}.$). Here $\tilde
X$ and $\tilde P$ are defined by a similar equation (\ref{Eq:X1}),
in which $\mu^{\ast}$ and $\omega^{\ast}$ are replaced,
respectively, by $\tilde {\mu}$ and $\tilde {\omega}.$ It is
interesting to notice that the spectrum of $\hat H_{free}$ is {\it
not} continuous, the interval of the spectrum is
$2\tilde{\omega}.$ For the case $\theta\to 0$ we have
$\tilde{\omega}\to 0,$ $2\tilde{\omega}\tilde A^\dagger \tilde
A\to \frac{1}{2\mu}p_i p_i.$ The Hamiltonian of a free particle in
commutative space is recovered. The zero-point energy $\tilde
E_0=\tilde{\omega}\hbar=\frac{1}{2\mu}\xi^2|\eta|$, which can also
be understood on the basis of the deformed momentum-momentum
noncommutativity. From Eq.~(\ref{Eq:xp}) it follows that the
deformed $\hat p-\hat p$ uncertainty relation reads
%%%%%%
$\Delta {\hat p_1} \Delta {\hat p_2}\ge \frac{1}{2}\xi^2|\eta|$.
%%%%%%
Taking $(\Delta \hat p_1)_{min}=(\Delta \hat
p_2)_{min}=\left(\frac{1}{2}\xi^2|\eta| \right)^{1/2},$ it follows
that the minimal energy $(\Delta \tilde E)_{min}$ corresponding to
$(\Delta {\hat p_i})_{min}$ is
%%%%%%
%\begin{equation}
%\label{Eq:dE}
$(\Delta {\tilde E})_{min}= \frac{1}{2\mu}\left[(\Delta \hat
p_1)_{min}^2+(\Delta \hat
p_2)_{min}^2\right]=\frac{1}{2\mu}\xi^2|\eta|$.
%\end{equation}
This shows that $(\Delta {\tilde E})_{min}=\tilde E_0.$ We
conclude that the deformed $\hat p-\hat p$ coherent state is
realized by the lowest energy state of a free particle.

In summary, in this paper first we clarify a subtle point related
to the equivalency between the deformed Heisenberg-Weyl algebra in
noncommutative space and the undeformed Heisenberg-Weyl algebra in
commutative space. For the case of both space-space and
momentum-momentum noncommuting, different from the case of only
space-space noncommuting, there is no singular-free linear
transformation to relate phase space variables between two
algebras, i.e. two algebras are not completely equivalent. It
follows that there is no well defined procedure to construct the
deformed position-position coherent state or the deformed
momentum-momentum coherent state from the undeformed
position-momentum coherent state. Then we demonstrate the
identification of the deformed position-position coherent state
with the lowest energy state of a cold Rydberg atom arranged in
appropriate electric and magnetic fields in the limit of vanishing
kinetic energy, and briefly show that the deformed
momentum-momentum coherent state is realized by the lowest energy
state of a free particle.

\vspace{0.4cm}

This work has been supported by the National Natural Science
Foundation of China under the grant number 10074014 and by the
Shanghai Education Development Foundation.

%%%%%%%%%%%%%%%%%%%%%%%%%%%%%%%%%%%%%%%%%%%%%%%%%%%%%%%%%%%%%%%%% saturate

\begin{flushleft}
a) E-mail: yinqijun@ecust.edu.cn

b) E-mail: jzzhang@ecust.edu.cn

$\ast)$ The corresponding author
\end{flushleft}
\clearpage

%%%%%%%%%%%%%%%%%%%%%%%%%%%%%%%%%%%%%%%%%%%%%%%%%%%%%%%%%%%%%%%%%%%%%%%%%%%

\end{document}